# An Environmental Watch System for the Andean countries: *El Observatorio Andino*

Ángel G. Muñoz[1,2], Patricio López[3], Ramón Velásquez[4], Luis Monterrey[4], Gloria León[5], Franklyn Ruiz[5], Cristina Recalde[6], Jaime Cadena[6], Raúl Mejía[6], Marcos Paredes[7], Juan Bazo[7], Carmen Reyes[7], Gualberto Carrasco[8], Yaruska Castellón[8], Claudia Villarroel[9], Juan Quintana[9], Avel Urdaneta[1]

*[1]Centro de Modelado Científico (CMC), La Universidad del Zulia, Maracaibo, Venezuela.*
*[2]Centro Internacional para la Investigación del Fenómeno El Niño (CIIFEN), Guayaquil, Ecuador.*
*[3]Agencia Estatal de Meteorología (AEMET), España*
*[4] Servicio de Meteorología de la Fuerza Aérea Venezolana (SEMETFAV), Venezuela.*
*[5]Instituto de Hidrología, Meteorología y Estudios Ambientales (IDEAM), Colombia.*
*[6]Instituto Nacional de Meteorología e Hidrología (INAMHI), Ecuador.*
*[7]Servicio Nacional de Meteorología e Hidrología (SENAMHI), Perú.*
*[8] Servicio Nacional de Meteorología e Hidrología (SENAMHI), Bolivia.*
*[9]Dirección Meteorológica de Chile (DMC), Chile.*

---

*Corresponding author:* Ángel G. Muñoz, Centro de Modelado Científico (CMC). La Universidad del Zulia, Maracaibo, 4004, Venezuela.
E-mail: agmunoz@cmc.org.ve

## Abstract

An experimental Environmental Watch System, the so-called Observatorio Andino-OA (Observatorio Andino), has been implemented in Venezuela, Colombia, Ecuador, Peru, Bolivia, and Chile over the past two years. The OA is a collaborative and regional network that aims to monitor several environmental variables and develop accurate forecasts based on different scientific tools. Its overall goal is to improve risk assessments, set up Early Warning Systems, support decision-making processes, and provide easily- and intuitively-understandable spatial maps to end-users. The initiative works under the scientific and logistic coordination of the Centro de Modelado Científico (CMC) at Zulia University, Venezuela, and the Centro Internacional para la Investigación del Fenómeno 'El Niño' (CIIFEN), and is operated at a local level by the National Weather Services (NWSs) of the aforementioned six Andean nations. The OA provides several freely-available model outputs including meteorological and hydrological forecasts, droughts, fire and flood indices, ecosystems dynamics, climate and health applications, and five-day high-resolution oceanographic predictions for the Eastern Pacific. This article briefly describes the current products, methodologies and dynamical and statistical modeling outputs provided by the OA. Also, a discussion on how these sets of tools have been put together as a coordinated, scientific watch and forecast system for each country and for the entire region is presented. Our experiences over the past two years suggest that this initiative would significantly improve the current decision-making processes in Andean countries.

**Capsule Summary**

A collaborative Environmental Watch System for Andean countries, the Observatorio Andino, is presented and discussed.

# 1. Introduction

The Andean countries are located in the western and northern regions of South America. Most of them lie in the tropical strip, where complex atmospheric dynamics and the lack of a comprehensive understanding of mesoscale and microphysical phenomena lead to difficult forecasting challenges. More than 7,000 km long, typically 200 km wide and about 4,000 m average height, the Andean Mountain Chain includes complex topographic features, with substantial repercussions on the meteorological and climatological behavior of the atmosphere in the region. The climate in this area shows large variability (Garreaud 2009, Garreaud et al. 2008) depending on altitude, location and proximity to the sea and to the Amazon region.

The ground observational network of the six Andean countries is neither homogeneously distributed nor dense enough and only a few National Weather Services (NWSs) have weather radars or radiosounding stations working on a regular basis. Satellite products are the most common and useful tools for nowcasting tasks in the region, while the Global Forecast System, GFS, (Kalnay et al., 1990) is the most frequently used model for operational forecasting.

Until 2006, efforts for building a regional and collaborative Climate Watch System for the Andean countries have been scarce. In 2007, CIIFEN received the approval and funding support from the Inter-American Development Bank (IDB) for the execution of a regional project entitled '*Climate Information applied to Agricultural Risk Management in the Andean countries'* (CIIFEN, 2007). Its implementation provided a valuable tool to the agricultural sector and, at the same time, allowed the development of a common methodology for the elaboration of agro-climatic risk maps on selected crops. An important component of the project involved capacity building of the participant NWSs on both probabilistic and dynamic downscaling basis. The CMC signed a collaborative agreement with CIIFEN that included support for the development of the necessary methodology, computational models, infrastructure, and training for running dynamical downscaling models as well as an experimental Seasonal Forecast System for the involved countries, especially running a General Circulation Model (GCM) in order to provide the necessary boundary conditions. This collaboration established the basis of the OA in 2008.

The OA main goal is to provide, in addition to the existing tools in each country, a collaborative network that ultimately increases the efficiency of decision-making processes, especially in terms of getting more accurate forecasts and exchanging experiences on data, information and scientific products, all of them with a standardized methodology and a web sharing service.

This paper is organized as follows: section 2 introduces the members of the OA and its products; section 3 describes the general forecast and verification methodologies; section 4 pinpoints various difficulties evidenced in the regional project; and section 5 discusses the achievements and future perspectives of the regional initiative.

## 2. The Observatorio Andino and Its Products

The aim of the OA is to monitor and forecast key environmental variables and develop accurate forecasts based on different scientific tools, in order to help decision makers to improve Risk Management and set up efficient Early Warning Systems. The OA provides several model outputs for meteorological, seasonal and hydrological forecasts, five-day high resolution oceanographic prediction for the Eastern Pacific, droughts, fire and flood indices, ecosystem dynamics (like duckweed/algae occurrence in Maracaibo Lake) and Climate and Health (e.g. Malaria) applications, among other products.

Currently, the CMC and the CIIFEN provide the logistic and scientific coordination of the OA. Six NWSs and several universities and research centers, including the International Research Institute for Climate and Society-IRI, Columbia University in the City of New York (USA), the San Andrés Major University (Bolivia), the Catholic University (Colombia), and the Chile University (Chile), actively participate in the initiative. The NWSs are in charge of validating and verifying the OA products and their use in each national Early Warning System, whereas universities and research centers produce scientific tools, methodologies and training modules that are implemented by the NWSs.

The OA is organized as interconnected Working Groups (WGs) or Axes, namely Meteorology, Climatology, Hydrology, Air Quality, Climate and Health, and Climate Change. Each WG has its own procedures and products, but all of them exchange information and results with other axes and countries if necessary.

The Wiki page http://mediawiki.cmc.org.ve, available in Spanish, contains information of the OA's philosophy and objectives, its members, and useful step-by-step instructions for installing, configuring and executing the available models. It also describes several methodologies such as computing models climatologies and anomalies, executing seasonal forecasting, performing objective analysis, using Geographical Information System (GIS) interfaces and verification processes, to mention just a few.

Advances and difficulties are shared and discussed through email lists and videoconferences using the open-source software Enabling Virtual Organizations (EVO). The purpose of the videoconferences is to discuss the ongoing projects, products and new methodologies, and provide special assistance on models and several other technical issues.

# 3. General Forecast Methodology and Verification

Still in an experimental stage, the Observatorio Andino follows a 3-level hierarchical model execution flux, with the same approach for both short-term and seasonal products.

Level I involves global scale analysis and/or General Circulation Model (GCM) outputs that are used as first guess or boundary/initial conditions for downscaling models (Level II). The downscaling products, in turn, provide the necessary information for Level III, which consists of microscale and/or tailored applications, like hydrological forecasts or Climate and Health Early Warning Systems.

A very important issue is the forecasting verification. Forecasts should guarantee consistency, quality and value (Mason and Stephenson, 2008) if they will help decision makers. However, even when the OA internally employs different standard metrics for the measurement of different forecast attributes, they are in general not well understood by the final user or decision maker. Thus, for validation and verification methodologies[1], an additional approach which generates easily and intuitively understandable spatial maps to end-users has been provided to all OA partners.

In order to facilitate the validation process, a spatial map for each time slice of interest (daily, monthly, seasonal) involving the differences and/or the anomalies correlation between model output and observations, is generated. For this purpose, the station data are objectively analyzed using the Cressman (1959) iterative scheme according to the model resolution and considering the scheme's elevation corrections due to the terrain complexity (accordingly, the final fields recognize the Andes presence). Thus, the differences or anomalies correlation are calculated over the gridded fields. Even when some problems occur with the objective analysis due to the low station density in the Andes region, in general the procedure results are satisfactory and, overall, useful and well understood by decision makers.

Other simple approach consists in constructing time series from the variables of interest, that include the local data (station) and the model output, after a bi-linear interpolation to the station coordinates have been done for the latter. In general, this kind of interpolation provides good results in most of the cases. However, due to the complex terrain, other approaches using digital elevation models are in course of experimentation.

The models running in the OA are depicted in Table 1. The main execution procedures are automated, as well as the post-processing schemes. In the next subsections a brief description of the general procedures is presented.

[1] Here we follow the convention that relates validation with the models *per se*, while verification is done for forecasts. At the OA, the NWSs are the institutions in charge of verifying products, while any institution could validate model outputs.

*a. Dynamical Weather Forecast*

At present, the OA provides 72-hour weather forecasts on a daily basis using the high resolution downscaling models MM5 (Michalakes, 2000) and WRF (Skamarock et. al, 2005). The GFS (Kalnay et. al, 1990) 3-hourly outputs and assimilation of SYNOP, METAR and TEMP reports are used as initial conditions. Each country has determined the best set of parameterizations for the models, which typically runs at 30 km or higher resolution.

The model outputs are considered valuable additional tools for the forecasting processes in countries where the Andes Mountain Chain provides complex disturbances that frequently GFS and other global models cannot resolve with enough skill.

*b. Dynamical Seasonal Forecast*

The NCAR Community Atmospheric Model version 3.1 (CAM3) (Collins et al., 2006) has been configured at T42L26 resolution in CMC for Atmospheric Model Intercomparison Project (AMIP) style runs with Green House Gases (GHGs) monthly variability from 1966 to present. The first 5 years have been discarded for spin-up reasons. The selected climatology corresponds to the 1971-2000 period. This particular configuration is hereby referenced as CAM_OA.

The present seasonal forecast methodology is sketched in Figure 1. On a monthly basis, the CAM_OA runs 6 ensemble members, where as tier-1: (a) two of them follow the persisted SST "e-folding" methodology (p_sst, see for example, Shuhua et al., 2008),

(b) two members use the SST forecast of the CFS model (cfs_sst, Saha et al. 2006), and (c) two realizations are obtained following the constructed analog (ca_sst, Van Den Dool, 1994) methodology. For all members the lead's monthly ice fraction coverage is described by the climatological values. For each member's output, the necessary initial and boundary conditions are extracted and written in the special (intermediate) format requested by the climatic versions of MM5 and WRF (CMM5 and CWRF, from now on), and are then available for the Andean NWSs through the OA web portal (http://ole2.org), which has been totally built with Open Source resources by CMC developers.

Each NWS downloads the needed files to execute the models in their own computational infrastructures, and, since January, 2010, using two different sets of physical parameterizations per model. Thus, a multiparametric multimodel ensemble is produced for each country, and then uploaded to the OA web portal for its publication after internal filters and discussion. In Figure 2 an OA's seasonal precipitation anomaly map is presented for all South America, with the corresponding observed rainfall anomaly for comparison.

These products are also used in each NWS for generation of agricultural risk maps and other products and tools for decision makers.

*c. Statistical Seasonal Forecast*

The IRI´s Climate Predictability Tool (CPT, http://iri.columbia.edu/outreach/software) is used to obtain probabilistic seasonal forecasts of mean, maximum and minimum temperature, as well as accumulated precipitation in the six Andean countries. Before this tool was implemented, the NWSs used to draw by hand on a map the forecasts based on the available information and subjective procedures. After CPT was introduced in the region in 2005-2006 (a CIIFEN-IRI collaboration), the main methodologies followed in the region consisted of a canonical correlation analysis or a principal component analysis, being the former most frequently used. The technical details are discussed by Mason and Baddour (2008). The usual predictors, downloaded every month from the IRI Data Library, are the observed sea surface temperature (SST) and, for both 850 mb and 500 mb levels, the geopotential height, zonal and meridional wind velocities, potential velocity and specific humidity. Local stations data for precipitation and temperature are included as predictands. The forecasts are provided in monthly, bi-monthly and seasonal formats, and each country dataset involves no less than 35 years of data. This process is done by the corresponding NWS through a regionalization procedure of its country, taking into account the principal component analysis and the forecaster's experience.

The forecast probabilities are firstly distributed as a nationwide spatial map by the NWSs and then send to CIIFEN for integration of all the products in regional wide maps. The final products appears monthly as temperature and rainfall anomalies charts in CIIFEN's Bulletin (http://www.ciifen-int.org).

*d. Oceanographic High Resolution Forecast*

The Regional Oceanic Modeling System (ROMS-AGRIF, Penven et al. 2007) has been configured for a computational domain in the Eastern Pacific. At present, the boundary and initial conditions are provided by the ECCO Consortium (Estimating the Circulation and Climate of the Ocean, Stammer et al. 1999) and the GFS (Kalnay et al., 1990). The OA runs daily ROMS for a 5-day high resolution (30 km) forecast, sharing in the web portal products like SST, surface salinity, vertical velocities (upwelling and downwelling) and marine currents. As an example, a vertical velocity map is shown in Figure 3. This kind of product is very useful as a basis for fisheries maps, providing approximate locations of nutrient-rich regions due to upwelling processes.

This OA's component has been developed by CMC in collaboration with the Comisión Permanente del Pacífico Sur (CPPS) in order to set up the same methodology developed by the NWSs but, in this case, for the Marine and Coastal Services of Colombia, Ecuador, Peru, and Chile.

*e. Dynamical Hydrological Forecast*

The Dynamical Hydrological Forecast (Level III) process is done in the OA by coupling the NOAH Land Surface Model (Schaake et al., 1996) with the Level II models, or directly using the latter forecasted precipitation, temperatures and wind outputs into the Variable Infiltration Capacity (VIC) Model of Liang et al. (1994, 2001). VIC is a macroscale (typical cell resolution > 1 km), semi-distributed hydrologic model that solves full water and energy balances. In the OA, the VIC is specifically configured for each basin of interest (the resolution depends on the selected basin) with the corresponding soil and vegetation type data.

For both procedures (coupled LSMs or uncoupled VIC Model), a bias correcting calibration procedure is applied to the raw output using historical, local streamflow data as reference (see Figure 4 for an example for one Ecuadorian basin).

After the calibration stage, the final outputs can be considered as a main tool for the corresponding Early Warning System in the involved countries.

*f. Other Applications*

Other applications include products related with droughts, floods, fires and ecosystem dynamics. In the case of droughts, Palmer (1965, 1968) indices are employed, while a composite map between runoff and hydrologic capacity of model cells are used to forecast possible floods. Likewise, the Chandler (1983) index is employed as a measure of joint probability of fire occurrence and propagation.

Climate and Health applications are focused mainly on malaria seasonal predictability for northwestern South America using the McDonald (1959) model. Given the necessary entomological and epidemiological parameters, the OA's high resolution outputs provide the climate information needed to run this epidemiological tool.

Finally, a new framework is related to Ecosystem Dynamics, especially Lemna (duckweed) population dynamics. In 2004 an important duckweed bloom took place in Maracaibo Lake (Tapias, 2010), the South American largest lake, bringing economic (e.g. fisheries) and health related (e.g. necrotic Lemna at lake shores produce an increase of diseases) problems to human populations in those coastal zones. Recently, the CMC provided an application known as CAVEL (Tapias, 2010) that makes use of MODIS VIS and IR data (Barnes et al. 2002) for providing normalized vegetation index (NDVI) maps, and time series of total surface coverage. As an example, Figure 5 shows a composite of Lemna images in visible and normalized difference vegetation index (NDVI).

# 3. Present Difficulties and Strategies

Although several difficulties arose in the process of the OA development, the efforts and decisions of the six NWSs were of fundamental importance to promote the OA initiative.

The Statistical and Dynamic Modeling Training sessions, organized by CIIFEN in the framework of the IDB project, have been a crucial component for providing the necessary human resources in the Andean region. These training processes have been composed by three workshops. The first of them, regarding Statistical Modeling and CPT, was held in Maracay (Venezuela) in October, 2007. The other two workshops, focused on Dynamic Downscaling Models installation, execution and post-processing, were held in Lima (Perú) in November, 2007 and in Guayaquil (Ecuador) in June, 2008. The OA has made extensive use of its wikipage, webinars and videoconferences for standardizing the methodologies, updating the different scripts and products, discussing the regional results and solving technical issues.

Despite the advances made since its creation the OA does not have its own funding and survives thanks to the support given by its members. Each NWS or institution provides available data, computers, communications and human resources, but there is clearly need for additional resources. The lack of personnel has forced the OA to automate most of the tasks, while the scarceness of computers dedicated to modeling has become an important motivation to share the resources of all the institutions in a new project called ANDESGrid. Certainly, the OA has transformed some disadvantages into advantages. But funding is still needed to guarantee the survival of this regional collaboration, mainly to prepare more human resources, facilitate exchange of experiences through regional meetings, and buy more computing and storage units.

Even when political differences between countries have not been an important barrier, more attention from local governments is definitely needed for supporting this initiative. For instance, in Venezuela there has been an important funding reduction for almost every public university, affecting severely the evolution of this project.

The perspectives of the OA not only aim to ensure local support but also build new partnerships with international institutions that may strengthen and extend the actual lines of action and products, as well as the related researching and development activities.

# 4. Concluding Remarks

In the last two years, many efforts have been directed to provide several scientific environmental tools for decision makers by both operative and research institutions in the Andean countries.

Among the OA's main achievements in this period, the most important is the enhancement of the NWSs' human resources by means of a continuous technical support for forecast, modeling, verification, objective analysis and many other issues, everything in their native language. Human resources have considerably increased their expertise on dynamical and statistical modeling, data processing and formats, model assimilation procedures, efficient use (and sharing) of computational resources, and even on the dynamical systems which modulates the climate on their countries, but with a more regional perspective and understanding than before.

The Project has also succeeded in standardizing forecast, data formats and verification/validation methodologies in the Andean countries, providing common models, tools and procedures which are used on a daily basis from Venezuela to Chile. The key tool has been the Wiki page, which contains all the needed steps for each task, and that by itself is one of the most important elements for ensuring the long-term continuity of the OA. Even being a technical reference, it counts today several thousand visits, and not only from the Andean countries.

With around 700 visits each month in the last two years, the OA product webpage is providing in one place different tools for helping decision makers on planning future actions and improving risk management in the Andean countries.

*Acknowledgments*

The Observatorio Andino acknowledge the invaluable contribution of several people at different countries (Estatio Gutiérrez, Rainer Schmitz, Daniel Ruiz, Marcos Andrade, Luis Blacutt, Mercy Borbor, Bryan Jordan, Joaquín Díaz, Xandre Chourio, Siulluz Reverol, Eugenio Tapias, Euselyne Sebrian) and Institutions (CMC, CIIFEN, SEMETFAV, IDEAM, INAMHI, SENAMHI Perú, SENAMHI Bolivia, DMC, IDB, FUNDACITE ZULIA, IVIC, AEMET, CPPS) that have been key actors for keeping the OA Project alive. Special thanks deserve the IRI's crew (Walter Baethgen, Lisa Goddard, Simon Mason, David DeWitt, Liqiang Sun, Tony Barnston, Gilma Mantilla, Esteban Andrade and others) for their scientific support, suggestions and training. The authors are grateful to Brant Liebmann and three anonymous referees for useful comments on the original version of the manuscript.

## List of Figures

FIG 1. Dynamical seasonal forecast methodology. A total of 6 realizations (top) are provided for all the participant institutions, which use the CAM_OA outputs for their own configurations of the dynamical downscaling models (middle). The final output (bottom) is an experimental multiparametric multimodel ensemble.

FIG. 2. South American seasonal anomaly precipitation forecast (above) and observations (below) for the period January-February-March, 2010. The forecast (CWRF model at 30

km resolution) corresponds to the ensemble downscaling product with bias correction. The observed field data is from the Climate Prediction Center.

FIG. 3. Upwelling (vertical velocity) map for the Eastern Pacific using ROMS (see section 3d).

FIG. 4. An example of observed and simulated (VIC) monthly streamflows ($m^3/s$ ) over the period from January, 2002 through December, 2004 for the Babahoyo basin in Ecuador (see section 3e). Observed data appear in red. Simulations correspond to two different values of the VIC infiltration parameter (Binf).

FIG. 5. An example of the ecosystem dynamic monitoring process for the Maracaibo Lake and Lemna population in visible (MODIS, top) and NDVI (CAVEL, bottom) for five months (from left to right: February, 2005; February, 2006; June, 2007; September, 2008; and April, 2009). See section 3f.

TABLE 1. Models currently working at the OA

| Model/index (version) | Hierarchy Level/Type | Executed by | Initial/Boundary Conditions | Domain | More Info in Section |
|---|---|---|---|---|---|
| GFS | I – Daily | NOAA | GDAS | Global | 5.a |
| CAM_OA (CAM-3.1) | I - Seasonal | CMC | p_SST, cfs_SST, ca_SST(2-tiered) | Global | 5.b |
| WRF (WRF-3.1) | II - Daily | CMC / Andean NWSs (6) | GFS, local data | South America /Country | 5.a |
| MM5 (MM5-3.7) | II - Daily | CMC / Andean NWSs (6) | GFS, local data | South America /Country | 5.a |
| CWRF (WRF-2.2) | II – Seasonal | CMC / Andean NWSs (6) | CAM_OA | South America /Country | 5.b |
| CMM5 (MM5-3.7) | II – Seasonal | CMC / Andean NWSs (6) | CAM_OA | South America /Country | 5.b |
| CPT (CPT-9.03) | II – Seasonal | Andean NWS (6) | Various from IRIDL | Andean Countries | 5.c |
| ROMS (ROMS-AGRIF) | II – Daily | CMC | ECCO, GFS | Eastern Pacific | 5.d |
| VIC (VIC-4.0.6) | III-Seasonal | CMC,INAMHI | CWRF,CMM5 | Country | 5.e |
| Droughts, Floods, Fires | III–Daily/Seasonal | CMC | CWRF,CMM5 | South America | 5.f |
| CAVEL | III-Seasonal | CMC | MODIS | State / Province | 5.f |
| Malaria (McDonald) | III–Seasonal | CMC, INAMHI | CWRF | South America /Ecuador | 5.f |

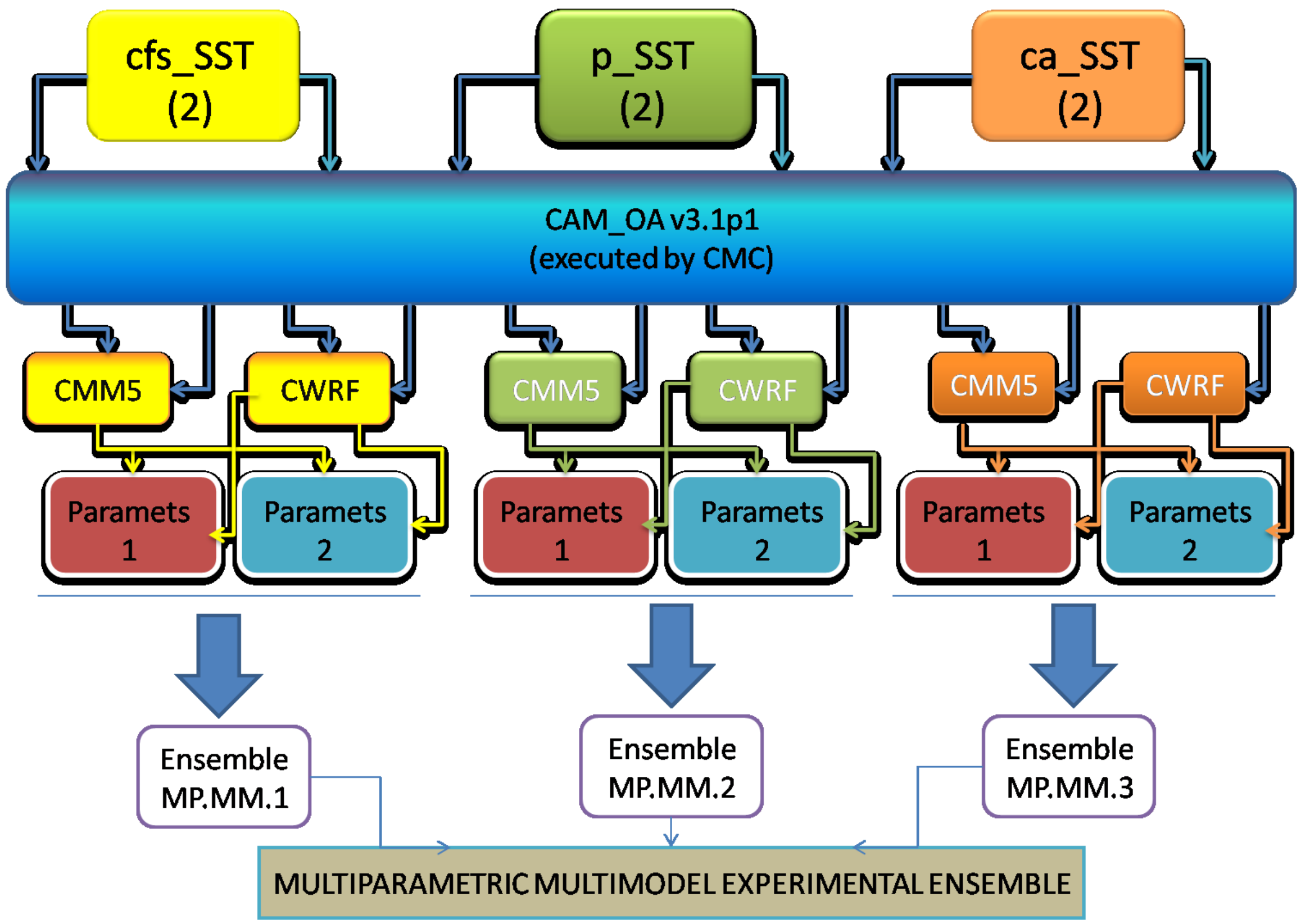

FIG 1. Dynamical seasonal forecast methodology. A total of 6 realizations (top) are provided for all the participant institutions, which use the CAM_OA outputs for their own configurations of the dynamical downscaling models (middle). The final output (bottom) is an experimental multiparametric multimodel ensemble.

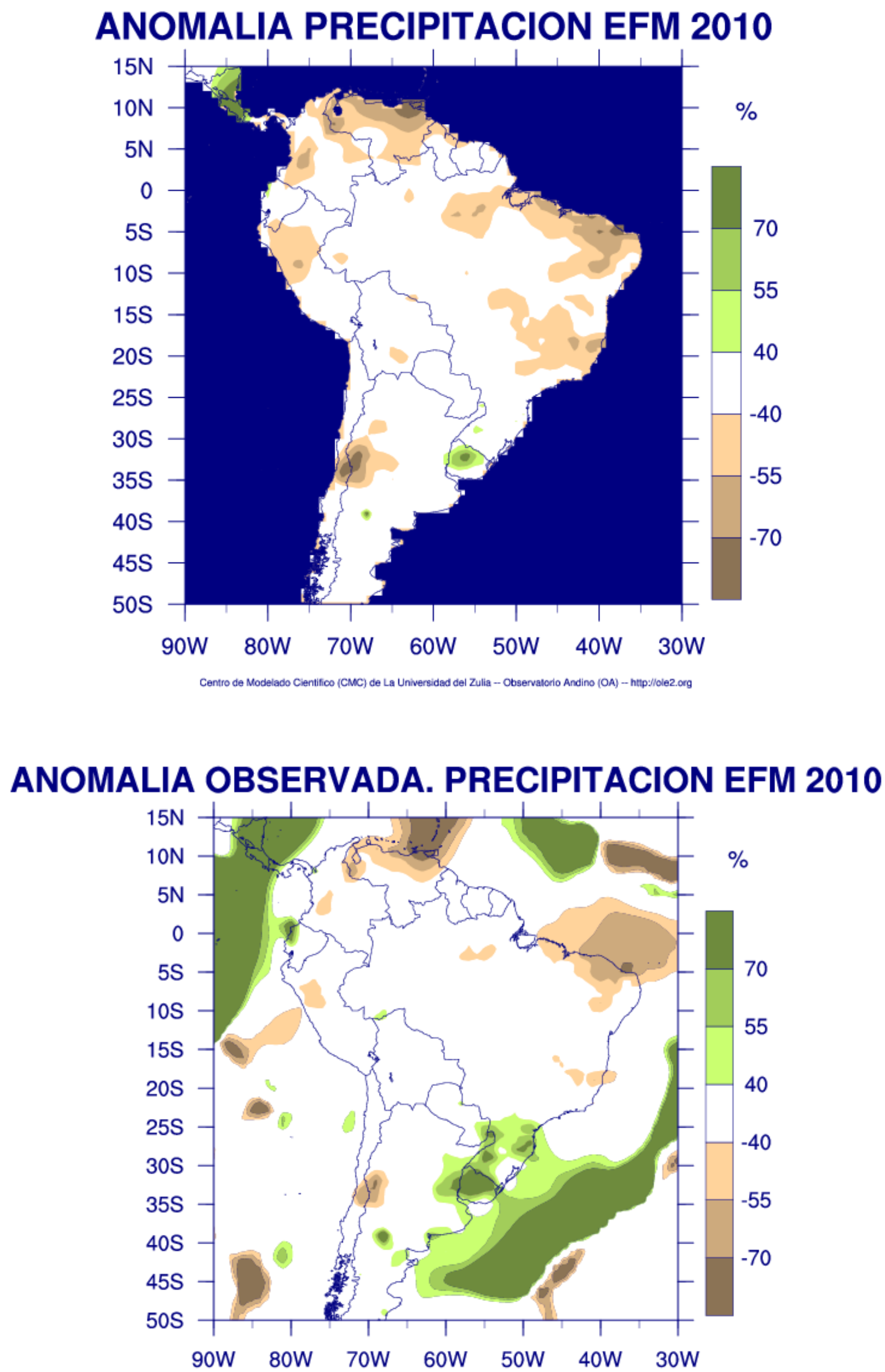

FIG. 2. South American seasonal anomaly precipitation forecast (above) and observations (below) for the period January-February-March, 2010. The forecast (CWRF model at 30 km resolution) corresponds to the ensemble downscaling product with bias correction. The observed field data is from the Climate Prediction Center.

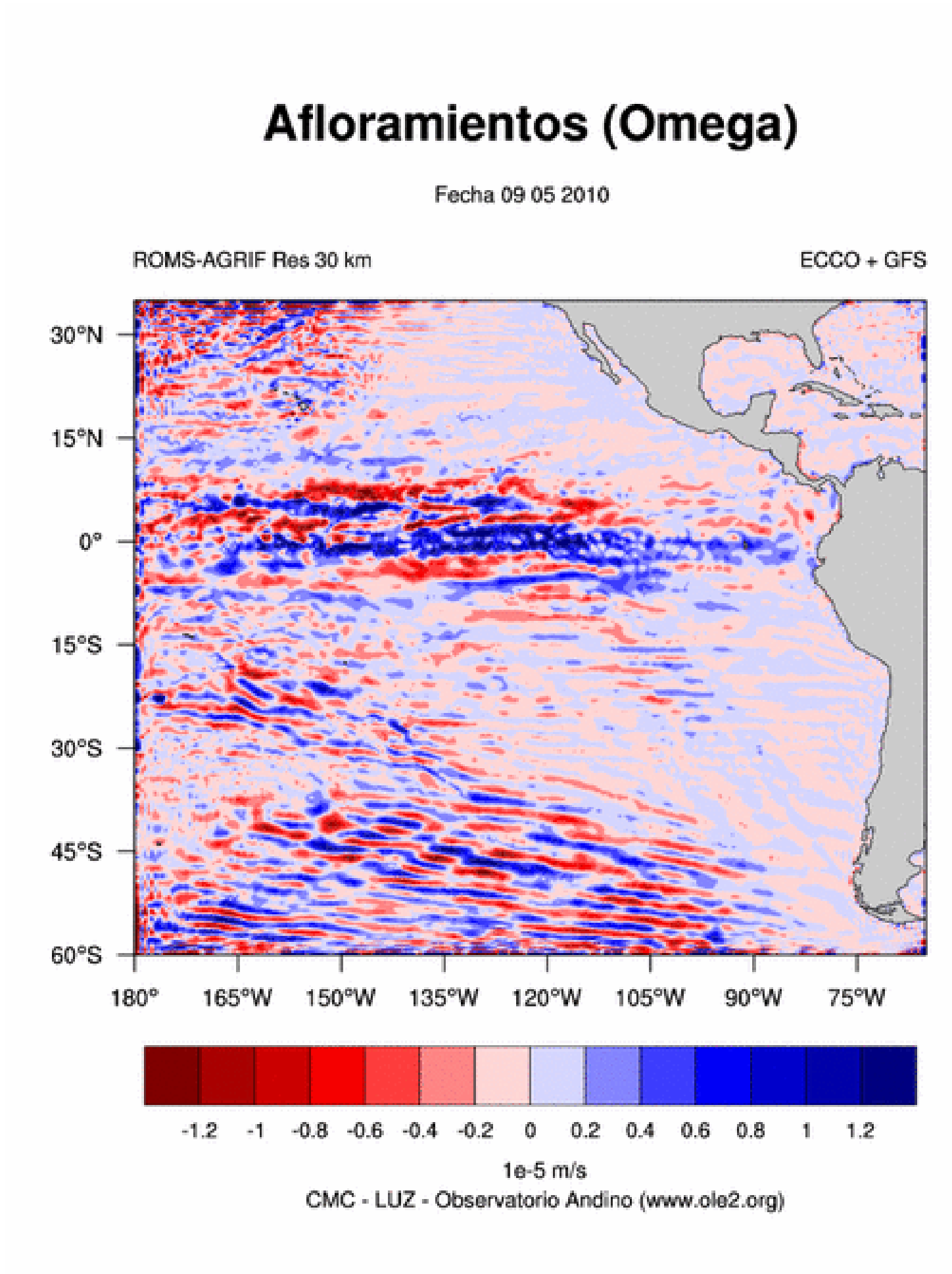

FIG. 3. Upwelling (vertical velocity) map for the Eastern Pacific using ROMS (see section 3d).

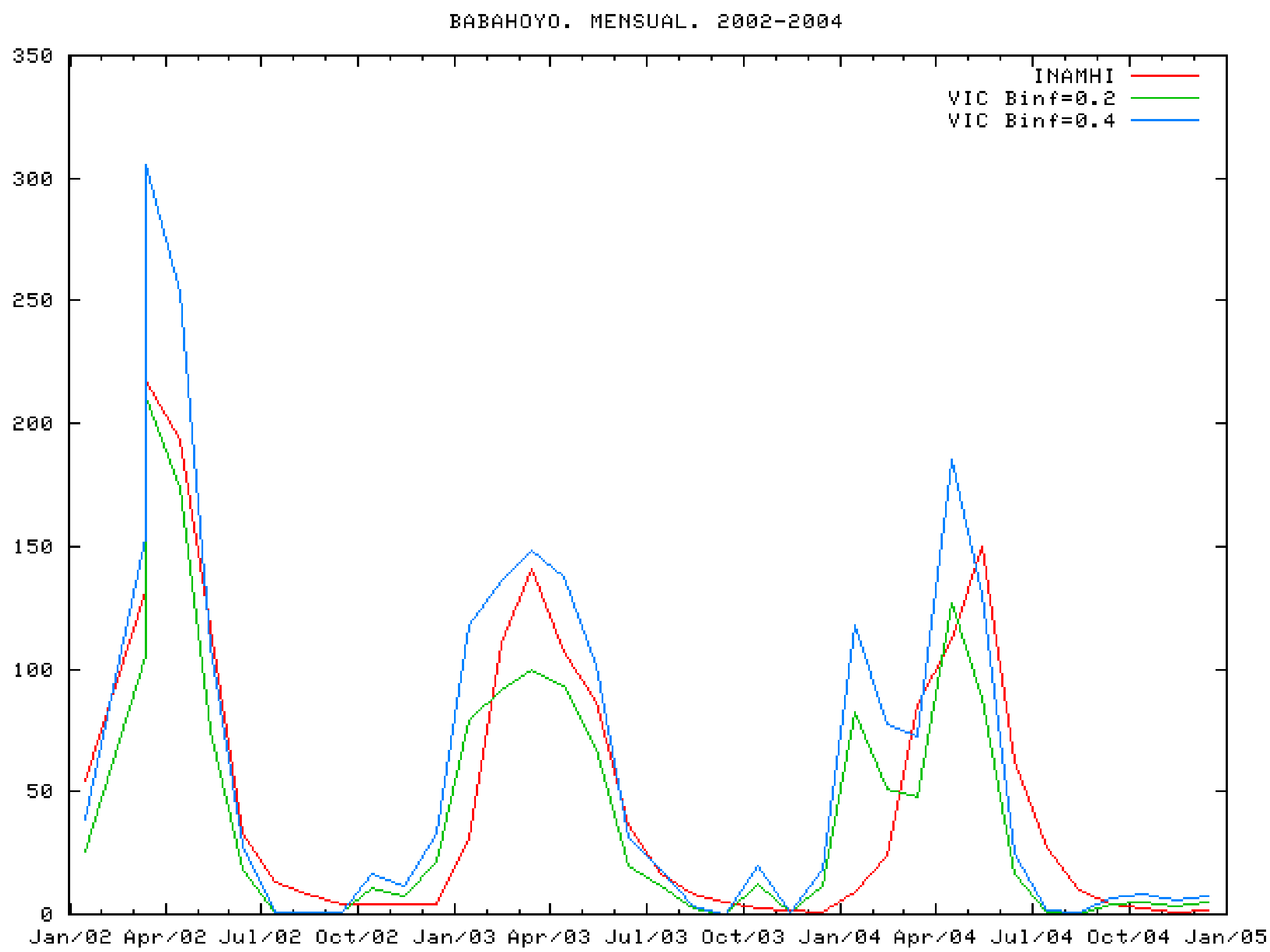

FIG. 4. An example of observed and simulated (VIC) monthly streamflows ($m^3/s$ ) over the period from January, 2002 through December, 2004 for the Babahoyo basin in Ecuador (see section 3e). Observed data appear in red. Simulations correspond to two different values of the VIC infiltration parameter (Binf).

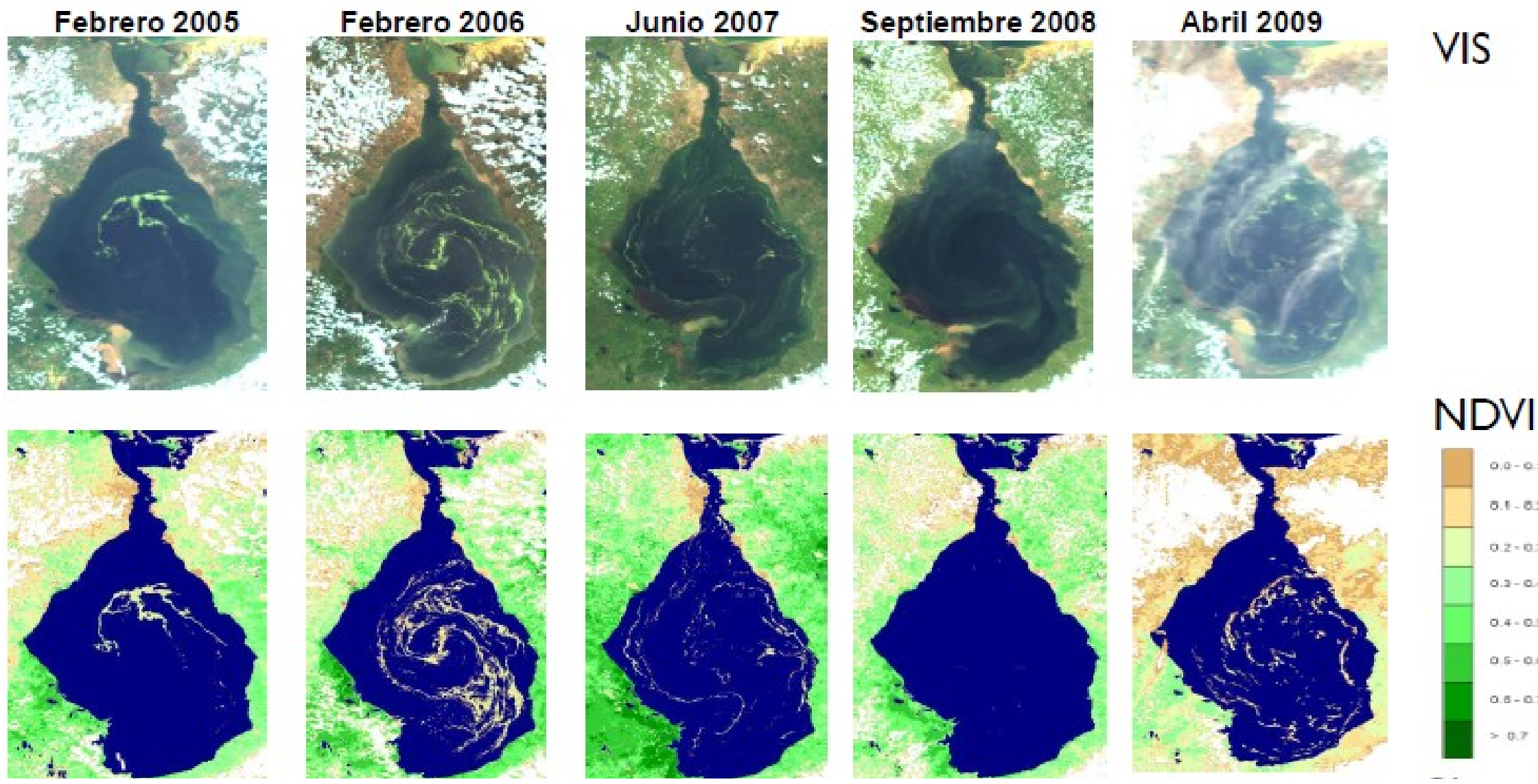

FIG. 5. An example of the ecosystem dynamic monitoring process for the Maracaibo Lake and Lemna population in visible (MODIS, top) and NDVI (CAVEL, bottom) for five months (from left to right: February, 2005; February, 2006; June, 2007; September, 2008; and April, 2009). See section 3f.